\documentclass[final,5p,times,sort&compress]{elsarticle}

\usepackage{graphicx}

\usepackage{amssymb}

\journal{Physics Letters B}

\begin{document}

\begin{frontmatter}



\title{Work distribution of an expanding gas and transverse energy
production in relativistic heavy ion collisions}


\author{Bin Zhang\corref{cor1}}
\ead{bzhang@astate.edu}
\cortext[cor1]{Corresponding author}

\author{Jay P. Mayfield}

\address{Department of Chemistry and Physics, Arkansas State University,
P.O. Box 419, State University, AR 72467-0419, USA}

\begin{abstract}

The work distribution of an expanding extreme relativistic gas
is shown to be a gamma distribution with a different shape
parameter as compared with its non-relativistic counterpart.
This implies that the shape of the transverse energy distribution 
in relativistic heavy ion collisions depends on the particle 
contents during the evolution of the hot and dense matter. 
Therefore, transverse energy fluctuations provide additional 
insights into the Quark-Gluon Plasma produced in these collisions.

\end{abstract}

\begin{keyword}

Work distribution \sep Transverse energy \sep Relativistic heavy
ion collisions


\end{keyword}

\end{frontmatter}


\section{Introduction}
\label{intro}


Transverse energy is an important characteristic of
relativistic heavy ion collisions aiming at creating conditions
similar to those existed in the early Universe 
\cite{Adcox:2001ry,Adams:2004cb,Adcox:2004mh,Adams:2005dq,
Collaboration:2011rta,Malek:2013saa}.
The hot and dense matter produced immediately after the 
collision of a nucleus going in the longitudinal
direction and the other in the opposite direction can
be considered as being composed of transverse fluid slices
undergoing longitudinal expansion following the two
receding nuclei \cite{Bjorken:1982qr}. In the nucleon-nucleon
center-of-mass frame, a slice that is closer to a nucleus
has a bigger longitudinal speed. Transverse energy
is longitudinally boost-invariant and therefore directly
reflects the condition of the local rest frame (slice) 
irrespective of its longitudinal flow speed. It is sensitive 
to the longitudinal work between adjacent slices and thus carries 
information about the evolution of the hot and dense
matter produced in heavy ion collisions \cite{Ruuskanen:1984wv,
Eskola:1992bd,Gyulassy:1997ib}.

Experimentally, the transverse energy distribution
is approximately a gamma distribution \cite{Chaudhuri:1993db}.
Interestingly, the work distribution for the adiabatic
compression or expansion of a dilute and interacting 
classical gas has been calculated by Crooks and Jarzynski 
\cite{Crooks:2007a}, and it is also a gamma distribution. 
The analogy between a longitudinally expanding Quark-Gluon 
Plasma and the adiabatic expansion of a classical gas in 
a cylinder prompts us to look at the latter more carefully.
Crooks and Jarzynski's calculation is for a non-relativistic
gas. Relativistic effects can be important for quarks and
gluons in a Quark-Gluon Plasma. In the following, we will 
show that for an extreme relativistic gas, the work 
distribution is also a gamma distribution. But the extreme
relativistic work distribution has a different shape
parameter relative to the non-relativistic one. The
transverse energy distributions can also be calculated,
and they are gamma distributions similar to the work
distributions. Because of the simplicity of the model,
the two parameters of the transverse energy distribution
have clear physical meanings. They are shown to reflect 
important properties of the evolution of the system.

\section{Work distribution and transverse energy production}
\label{workd}


The work distribution for a non-relativistic, dilute, interacting,
classical gas undergoing adiabatic compression or expansion
is given by \cite{Crooks:2007a}
\begin{equation}
\label{equ:wds1}
\rho(W)=\frac{\beta}{|\alpha|\Gamma(k)}
\left(\frac{\beta W}{\alpha}\right)^{k-1}
\exp\left(-\frac{\beta W}{\alpha}\right)\theta(\alpha W).
\end{equation}

Here $W$ is the work on the system. It is positive for compression
and negative for expansion. In three dimensions, $\alpha$ 
is related to the initial volume $V_0$ and final volume $V_1$
by
\begin{equation}
\label{equ:wds2}
\alpha=\left(\frac{V_0}{V_1}\right)^{2/3}-1.
\end{equation}
$\alpha$ is positive for compression and negative for expansion.
The unit step function $\theta(\cdot)$ in Eq.~(\ref{equ:wds1}) 
ensures that $W$ and $\alpha$ always have the same sign.
$\beta$ is the inverse of the initial fundamental temperature. 
In three dimensions,
$k$ is related to the total number of particles $N$ via $k=3N/2$.
Unless stated otherwise, we will use the natural unit system, 
in which the reduced Planck constant $\hbar$, the speed of light 
in vacuum $c$, and the Boltzmann constant $k_B$ are set to $1$.

The distribution of the magnitude of work then acquires the form
\begin{equation}
\label{equ:wds3}
\bar{\rho}(|W|)=\frac{\beta}{|\alpha|\Gamma(k)}
\left(\frac{\beta |W|}{|\alpha|}\right)^{k-1}
\exp\left(-\frac{\beta |W|}{|\alpha|}\right).
\end{equation}
It is a gamma distribution described by shape $k=3N/2$ and scale
$s=|\alpha|/\beta$.

In the following, we will derive the work distribution for
the extreme relativistic case and make some comparisons.
A good starting point is the number of energy states
with the energy of the gas less than $E$. It can be described
by the asymptotic formula \cite{Pathria:1996a}
\begin{equation}
\label{equ:wds4}
\Phi(E;V)=\frac{1}{(2\pi)^{3N}}\frac{V^N}{N!}
\frac{(8\pi)^NE^{3N}}{(3N)!}=\frac{V^N}{\pi^{2N}N!}\frac{E^{3N}}{(3N)!}.
\end{equation}
The density of states can now be calculated as
\begin{equation}
\label{equ:wds5}
g(E;V)=\frac{\partial \Phi}{\partial E}=
\frac{V^N}{\pi^{2N}N!}\frac{E^{3N-1}}{(3N-1)!}.
\end{equation}
This leads to the partition function
\begin{equation}
\label{equ:wds6}
Z(\beta,V)=\int dE g(E;V) \exp(-\beta E)=
\frac{V^N}{\pi^{2N}N!}\frac{1}{\beta^{3N}}.
\end{equation}
Energy $E$ follows the canonical distribution
\begin{eqnarray}
\label{equ:wds7}
P(E;\beta)&=&\frac{g(E;V)}{Z(\beta,V)}\exp(-\beta E) \nonumber \\
 &=&
\beta\frac{(\beta E)^{3N-1}}{(3N-1)!}\exp(-\beta E).
\end{eqnarray}
The work during an adiabatic process is the 
change in the internal energy, i.e.,
\begin{equation}
\label{equ:wds8}
W=E_1-E_0.
\end{equation}
Assuming ergodicity, $\Phi(E;V)$ is an adiabatic invariant, and 
$E_1$ can be related to $E_0$ via
\begin{equation}
\label{equ:wds9}
E_1=\left(\frac{V_0}{V_1}\right)^{1/3}E_0.
\end{equation}
Therefore, 
\begin{equation}
\label{equ:wds10}
W=\left(\left(\frac{V_0}{V_1}\right)^{1/3}-1\right)E_0=\alpha E_0.
\end{equation}
Here 
\begin{equation}
\label{equ:wds11}
\alpha=\left(\frac{V_0}{V_1}\right)^{1/3}-1
\end{equation}
is different from the non-relativistic formula given in 
Eq.~(\ref{equ:wds2}).
Now the work distribution is given by
\begin{eqnarray}
\label{equ:wds12}
\rho(W)&=&\int dE_0 P(E_0;\beta) \delta(W-\alpha E_0)\nonumber\\
&=&
\frac{\beta}{|\alpha|\Gamma(k)}
\left(\frac{\beta W}{\alpha}\right)^{k-1}
\exp\left(-\frac{\beta W}{\alpha}\right)\theta(\alpha W),
\end{eqnarray}
where $k=3N$ and $\alpha$ is given by Eq.~(\ref{equ:wds11}).
The distribution of the magnitude of work is also given
by Eq.~(\ref{equ:wds3}) with $k$ and $\alpha$ given by
the extreme relativistic formulas above. Therefore, the
extreme relativistic case has the same work distribution
compared to the non-relativistic case, but it has different 
formulas for the parameters. In particular, the shape parameter 
changes from $k=3N/2$ in the non-relativistic case to $k=3N$
in the extreme relativistic case.

Making use of the expression for the free energy
\begin{equation}
\label{equ:wds13}
F(\beta,V)=-\frac{1}{\beta}\ln Z(\beta,V)=
-\frac{1}{\beta}\ln\left(\frac{V^N}{\pi^{2N}N!}
\frac{1}{\beta^{3N}}\right),
\end{equation}
the work distribution can be shown to satisfy
the Jarzynski equality \cite{Jarzynski:1997a,Jarzynski:1997b}
\begin{eqnarray}
\label{equ:wds14}
\lefteqn{-\ln\langle\exp(-\beta W)\rangle
=-\ln\int dW\rho(W)\exp(-\beta W)}  \nonumber\\
&=N\ln\left(\frac{V_0}{V_1}\right)=\beta\Delta F,
\end{eqnarray}
where $\Delta F=F(\beta,V_1)-F(\beta,V_0)$ is the change
in the free energy between two states with the same $\beta$
and different volumes.

It is also straightforward to show that the work distribution
for the forward process $\rho_F$ and that for the corresponding 
reverse process $\rho_R$
satisfy the Crooks fluctuation theorem \cite{Crooks:1999a,Crooks:2000a}
\begin{equation}
\label{equ:wds15}
\frac{\rho_F(W)}{\rho_R(-W)}=
\left(\frac{V_1}{V_0}\right)^N\exp(\beta W)=
\exp(\beta(W-\Delta F)).
\end{equation}

\begin{figure}[htb]
\includegraphics[width=224pt,bb=100 45 385 305,clip]{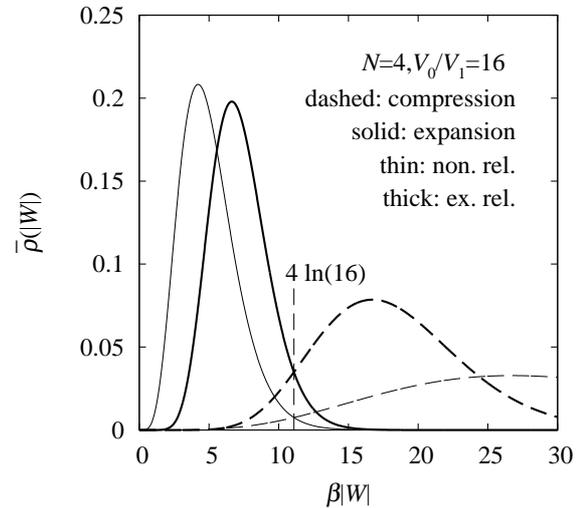}
\caption{Work magnitude distributions for the compression and expansion of
a non-relativistic (non. rel.) gas and an extreme relativistic (ex. rel.) gas.}
\label{fig:cmp1}
\end{figure}

Now let us compare the non-relativistic case and the extreme relativistic
case. Fig.~\ref{fig:cmp1} gives the work magnitude distributions. In both
the non-relativistic case and the extreme relativistic case, the 
compression curve and the expansion curve meet at 
$\beta |\Delta F|=4\ln(16)$. The extreme relativistic curves
are closer to $4\ln(16)$ than the corresponding non-relativistic
curves. Consequently, the extreme relativistic case has
higher probability of having $W<\Delta F$. However, the second
law of thermodynamics, i.e., the average work $\langle W\rangle$
is not smaller than $\Delta F$, is still valid 
\cite{Jarzynski:2001a}. From the fact
that the average of a gamma distribution is the product of 
the shape and scale parameters, for the compression case,
the non-relativistic average work $\langle W\rangle_n=
3N/(2\beta)((V_0/V_1)^{2/3}-1)$,
and the extreme relativistic average work $\langle W\rangle_e=
3N/\beta((V_0/V_1)^{1/3}-1)$.
This can be compared with $\Delta F = N/\beta\ln(V_0/V_1)$.
We arrive at the relation 
$\langle W\rangle_n>\langle W\rangle_e>\Delta F>0$.
It shows the ordering of $\langle W\rangle_n$ and $\langle W\rangle_e$,
and that both the non-relativistic and extreme relativistic cases
satisfy the second law of thermodynamics. Likewise, for the expansion
case, $0>\langle W\rangle_n>\langle W\rangle_e>\Delta F$.

\begin{figure}[htb]
\includegraphics[width=224pt,bb=100 45 385 305,clip]{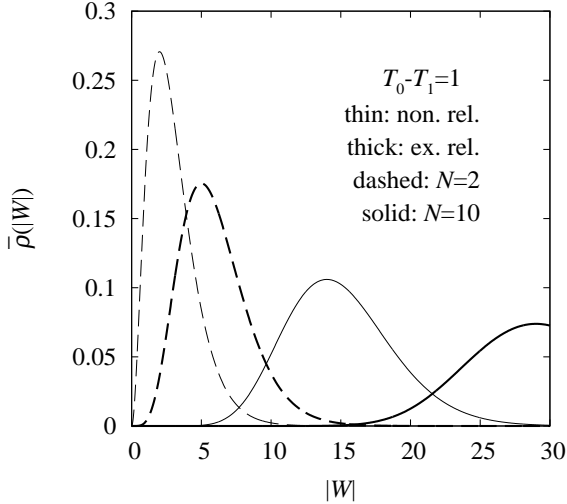}
\caption{Work magnitude distributions for a non-relativistic 
gas and an extreme relativistic gas undergoing the same 
temperature change.}
\label{fig:cmp2}
\end{figure}

With the same compression or expansion ratio $V_0/V_1$, the
non-relativistic and the extreme relativistic cases have different
$\alpha$ values as given by Eqs.~(\ref{equ:wds2}) and (\ref{equ:wds11}).
Thus the work magnitude distributions have different scale 
parameters $s=|\alpha|/\beta$. During an adiabatic process,
the ensemble remains canonical. Therefore, for the non-relativistic
case with initial temperature $T_0$ and final temperature $T_1$,
\begin{equation}
\label{equ:wds16}
T_1=\left(\frac{V_0}{V_1}\right)^{2/3}T_0,
\end{equation}
and for the extreme relativistic case,
\begin{equation}
\label{equ:wds17}
T_1=\left(\frac{V_0}{V_1}\right)^{1/3}T_0.
\end{equation}
This leads to an interesting expression for the scale parameter,
i.e., $s=|T_1-T_0|=|\Delta T|$. This can be more relevant for
relativistic heavy ion collisions, where the hot and dense
matter can be considered as starting at some initial temperature
and stopping at some freeze-out temperature. Fig.~\ref{fig:cmp2} 
shows that non-relativistic and extreme relativistic gases have 
very different work distributions. Since they have the same
scale parameter, the difference in the means comes from different
shape parameters, and the extreme relativistic one is about twice
that of the non-relativistic case.

Before calculating the transverse energy distribution, we
will look at the final energy distribution. The final energy
can be related to the initial energy by $E_1=(V_0/V_1)^{\{2/3,1/3\}}
=qE_0$. Unless stated otherwise, the first choice in the 
braces is for a non-relativistic gas, and the second is for the 
extreme relativistic one. Now the final energy distribution
\begin{eqnarray}
\label{equ:wds18}
p(E_1)&=&\int dE_0P(E_0;\beta)\delta(E_1-qE_0) \nonumber \\
&=&\frac{1}{q}P\left(\frac{E_1}{q};\beta\right)=
P\left(E_1;\frac{\beta}{q}\right)
\end{eqnarray}
is a gamma distribution with shape $k=\{3N/2,3N\}$ and scale
$s=q/\beta=T_0(V_0/V_1)^{\{2/3,1/3\}}=T_1$. This is expected
as the ensemble remains canonical during the adiabatic process.

In order to get the transverse energy distribution, we need
to approximate the sum over particles by an integral.
For the non-relativistic case, 
\begin{eqnarray}
\label{equ:wds19}
\lefteqn{E_1=\sum_{i=1}^NE_{1,i}
=\int d^3p C \exp\left(-\frac{p^2}{2mT_1}\right)\frac{p^2}{2m}}\nonumber\\
\lefteqn{=C\int_0^\infty dp p^2 \exp\left(-\frac{p^2}{2mT_1}\right)
\frac{p^2}{2m}
\int_{-1}^1d\cos\theta\int_0^{2\pi}d\phi}\nonumber\\
\lefteqn{=D\int_{-1}^1d\cos\theta=2D.}
\end{eqnarray}
In the above, $E_{1,i}$ is the final energy of particle $i$.
$C$ and $D$ are constants. $\theta$ and $\phi$ are the polar and
azimuthal angles in the spherical coordinate system where the 
polar axis goes along the longitudinal direction.
The final transverse energy
\begin{eqnarray}
\label{equ:wds20}
\lefteqn{E_{1\perp}=\sum_{i=1}^NE_{1\perp,i}
=\int d^3p C \exp\left(-\frac{p^2}{2mT_1}\right)
\frac{p_\perp^2}{2m}}\nonumber\\
\lefteqn{=C\int_0^\infty dp p^2 \exp\left(-\frac{p^2}{2mT_1}\right)
\frac{p^2}{2m}
\int_{-1}^1\sin^2\theta d\cos\theta\int_0^{2\pi}d\phi}\nonumber\\
\lefteqn{=D\int_{-1}^1\sin^2\theta d\cos\theta=\frac{4}{3}D.}
\end{eqnarray}
Therefore, $E_{1\perp}=\frac{2}{3}E_1$.

For the extreme relativistic case,
\begin{eqnarray}
\label{equ:wds21}
\lefteqn{E_1=\sum_{i=1}^NE_{1,i}
=\int d^3p C' \exp\left(-\frac{p}{T_1}\right)p}   \nonumber\\
\lefteqn{=D'\int_{-1}^1d\cos\theta=2D',}
\end{eqnarray}
where $C'$ and $D'$ are constants.
\begin{eqnarray}
\label{equ:wds22}
\lefteqn{E_{1\perp}=\sum_{i=1}^NE_{1\perp,i}
=\int d^3p C' \exp\left(-\frac{p}{T_1}\right)p_\perp}\nonumber\\
\lefteqn{=D'\int_{-1}^1\sin\theta d\cos\theta=\frac{\pi}{2}D'.}
\end{eqnarray}
Hence, $E_{1\perp}=\frac{\pi}{4}E_1$.

The non-relativistic and the extreme relativistic cases can be
summarized into one formula $E_{1\perp}=aE_1$, where
$a=\{2/3,\pi/4\}$.

Now the final transverse energy distribution
\begin{eqnarray}
\label{equ:wds23}
d(E_{1\perp})&=&\int dE_1 p(E_1)\delta(E_{1\perp}-aE_1)
=\frac{1}{a}p\left(\frac{E_{1\perp}}{a}\right)\nonumber\\
&=&\frac{1}{aq}P\left(\frac{E_{1\perp}}{aq};\beta\right)
=P\left(E_{1\perp};\frac{\beta}{aq}\right)
\end{eqnarray}
is a gamma distribution with shape $k=\{3N/2,3N\}$ and
scale $s=(aq)/\beta=\{2/3,\pi/4\}T_1$. This tells us that
the shape parameter is very sensitive to the particle
contents during the evolution while the scale parameter
is slightly sensitive and is mainly determined by
the final temperature.

\section{Summary and discussions}
\label{summa}


The work distribution for an extreme relativistic gas
undergoing an adiabatic process is shown to be a gamma distribution
with a shape parameter twice as large as that for the
non-relativistic gas. Both cases have a scale parameter
that can be related to the change in the system temperature.
The corresponding transverse energy distributions are
also gamma distributions. In both the extreme relativistic
and the non-relativistic cases, the shape parameter
is the same as that for the work distribution, and
the scale parameter is related to the final temperature.
This gives insights into relativistic heavy ion collisions 
where transverse energy distributions are approximate
gamma distributions. In particular, a change in 
the particle mass can lead to a change in the shape
parameter, and the scale parameter is mainly determined
by the freeze-out temperature and depends slightly on
the particle mass. This may lead to deviations from
the superposition of nucleon-nucleon collisions as
has been calculated in \cite{Chaudhuri:1993db}. 
On the other hand,
the transverse energy distribution in proton-proton collisions
is also approximately a gamma distribution 
\cite{Chaudhuri:1993db}. Thus the parameters may reflect
details of proton-proton collisions.

In general, the density of states cannot be written
as a power of the energy. In particular, \ref{appe}
shows that
\begin{eqnarray}
\label{equ:wds24}
\lefteqn{g(E;V)=\frac{V^N}{(2\pi)^{3N}N!}\frac{(4\pi m^3)^N}{(N-1)!m}
\prod_{j=1}^N\int_0^{t^*}dt_j \cosh(2t_j)}\nonumber\\
\lefteqn{\left(\frac{E}{m}+N-\sum_{k=1}^N\cosh t_k\right)^{N-1}
\theta\left(\frac{E}{m}+N-\sum_{l=1}^N\cosh t_l\right),}
\end{eqnarray}
where
\begin{equation}
\label{equ:wds25}
t^*=\ln\left(\frac{E}{m}+1+\sqrt{\left(\frac{E}{m}+1\right)^2-1}\right).
\end{equation}
The energy distribution can be calculated numerically by making use of
Eq.~(\ref{equ:wds24}), but it 
is not a simple gamma distribution, and 
it is unlikely that the work distribution can be expressed 
as a gamma distribution.
It is possible that the work distribution can be approximated
by a gamma distribution interpolating between the
non-relativistic case and the extreme relativistic case.
If so, qualitative changes due to the change of
particle mass are also expected to show up strongly
in the effective shape parameter.

In addition to the finite mass corrections, other factors
come into play in relativistic heavy ion collisions. The
longitudinally expanding gas with local thermal equilibrium
can only give some general guidance for the effects of
final state interactions. Even for the extreme relativistic
case, as the kinetic energy is much larger than the rest
mass energy, particle production and annihilation can 
happen, and entropy is expected to change accordingly.
At this moment, the role of particle number changing
processes is still under intense investigation 
\cite{Xu:2007ns,Chen:2009sm,Zhang:2010fx,Chen:2010xk,
Zhang:2012vi,Fochler:2013epa,Huang:2013lia,Zhang:2013ota}.
As particle production during the expansion process leads
to more cooling, the entropy increase is expected to
be small \cite{Letessier:1993qa}. It is not clear whether
and by how much particle number changing processes affect
the proportion relationship between the initial and final
energies. If the relation is significantly modified,
there could be large deviations from the gamma distribution
for the transverse energy distribution.
Other than the particle number changing processes,
the changing particle contents, 
the transverse expansion, and the differential freeze-out
all contribute to the evolution of the transverse
energy distribution. Studies with dynamical
models will be necessary to sort out the details.

The above derivation of the transverse energy distribution
depends on local thermal equilibrium. Deviations from equilibrium
can lead to deviations from the gamma distribution. For example,
initial conditions based on the Glauber model \cite{Miller:2007ri}
or various saturation models \cite{Drescher:2006ca,Drescher:2007ax} 
can be very different from thermal initial conditions. The
difference may lead to observable deviations from the gamma 
distribution when higher order moments are studied. 
In other words, longitudinal flow results can complement
the widely investigated transverse flow analyses \cite{Heinz:2013th}.
During the late stage of the evolution, the viscosity
is large, and the system cannot maintain equilibrium. The
longitudinal work will be significantly reduced. At the other
extreme from equilibrium, no longitudinal work is expected for
the free streaming case, and the
transverse energy distribution does not change. Therefore,
as a first approximation, the change of the transverse energy
distribution due to the late stage can be neglected. To be a 
little more precise, if the average kinetic energy can be used as
a measure of the temperature in the non-equilibrium case, the
freeze-out temperature that goes into the transverse energy
distribution estimate should be higher than the non-equilibrium
``temperature". 

The gamma distribution is also limited to a classical gas.
Recently, there are some discussions of the possibility of 
forming a Bose-Einstein condensate in the early stage
of a relativistic heavy ion collision 
\cite{Blaizot:2011xf,Blaizot:2013lga}. If a Bose-Einstein
condensate is formed, the transverse energy distribution
may have some noticeable difference from a gamma distribution.
One can examine the key elements in the derivation of
the gamma distribution and see the difference. One thing
that does not change in the derivation is the relation between
the work during an adiabatic process and the initial energy.
The reason is that the relation between the energy of a single 
particle in a box and the volume is independent of whether
the particle is a Boson or a Fermion. If interactions
are negligible, the relation between the total energy of
the system and the volume is independent of whether the 
particles are Bosons or Fermions. However, the number of 
states with energy smaller than a given value depends 
on the number of ways to partition an integer 
\cite{Auluck1946a,Agarwala1951a}. Thus the density
of states depends on whether the system is composed of Bosons
or Fermions. Only in the non-degenerate limit does the work
distribution become a gamma distribution.

It is interesting to see what experimental data can teach us.
Our focus will be on relativistic heavy ion collisions where
hydrodynamics is successful in describing various
experimental observations. In particular, preliminary data from 
the PHENIX collaboration at the Relativistic Heavy Ion 
Collider \cite{Armendariz:2007zz} will be used. 
The transverse energy here is the transverse
energy measured in the electromagnetic calorimeter.
If $\mu$ is the average transverse energy, and $\sigma^2$ is
the variance, they can be related to the shape parameter $k$
and the scale parameter $s$ via $\sigma/\mu=1/\sqrt{k}$ and
$\sigma^2/\mu=s$. For Au+Au collisions at the nucleon-nucleon
center-of-mass frame energy $\sqrt{s_{\mathrm{NN}}}=200$ GeV, the 
0--5\% centrality bin has $\sigma/\mu=0.116$ and 
$\sigma^2/\mu=1.60$ GeV. Therefore, the shape parameter 
$k=74$, and the scale parameter $s=1.60$ GeV. 
They appear to be outside the ranges
expected from the commonly accepted initial particle number
and final temperature values. However, each centrality bin has a 
distribution of particle numbers \cite{Adler:2004zn}. 
Selection according to the number of particles may 
improve on the situation. 

\begin{figure}[htb]
\includegraphics[width=224pt,bb=100 55 370 305,clip]{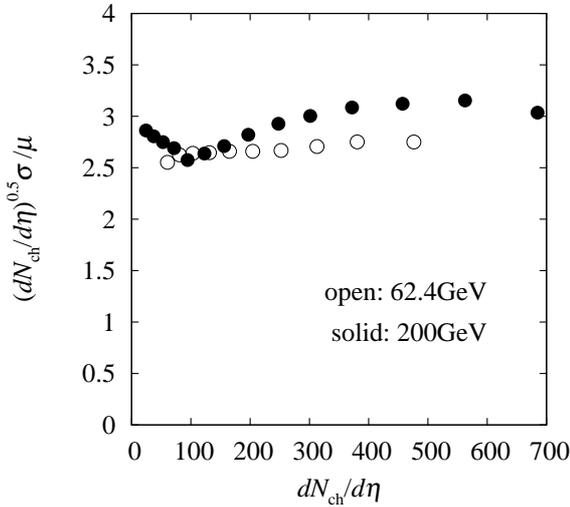}
\caption{Centrality dependence of the ratio of the relative
width $\sigma/\mu$ to $1/\sqrt{dN_{\mathrm{ch}}/d\eta}$ for
$\sqrt{s_{\mathrm{NN}}}=62.4$ GeV and $200$ GeV.
}
\label{fig:rel_wid2}
\end{figure}

The experimental data may provide more information. If using
the average number of particles for a centrality bin only 
introduces a common rescaling factor for all centralities, 
and if the particle contents remain the same over different
centralities, the ratio of $\sigma/\mu$ to $1/\sqrt{N}$
is expected to be independent of centrality. Assuming the charge
particle pseudo-rapidity distribution $dN_{\mathrm{ch}}/d\eta$ 
is $2/3$ of the particle pseudo-rapidity distribution, 
the centrality dependence of 
$\sigma/\mu/(1/\sqrt{dN_{\mathrm{ch}}/d\eta})$ should reflect
the particle contents. Fig.~\ref{fig:rel_wid2} shows this ratio
for $\sqrt{s_{\mathrm{NN}}}=62.4$ GeV and $200$ GeV. 
If other complications are not important, for the 200 GeV case,
as $dN_{\mathrm{ch}}/d\eta$ increases toward $100$, 
the particles responsible for the longitudinal work become 
lighter (more relativistic). The particles become
heavier (less relativistic) as $dN_{\mathrm{ch}}/d\eta$ 
increases further from 100. One should bear in mind that the error 
bars are on the order of 5--10\%, and there can be other 
factors affecting the centrality dependence. It also appears 
that 200 GeV has heavier particles for many centrality bins
compared with the 62.4 GeV case.

In a recent preprint \cite{Adler:2013aqf}, 
the PHENIX collaboration compared the experimental
transverse energy distributions to results from some
models based on the superposition of nucleon-nucleon collisions.
They demonstrated that the experimental data favor 
the number-of-constituent-quark-participant model. 
However, the superposition of nucleon-nucleon 
collisions cannot generate collective flow (longitudinal, radial,
or elliptic). As longitudinal flow reduces the event transverse 
energy, even the number-of-constituent-quark-participant
model can give very different transverse energy distributions
when the hydrodynamic evolution of the hot and dense matter
is taken into account. On the other hand,
if no dynamical models can give satisfactory explanation
of the measured transverse energy distributions, it will be a
real challenge to reconcile different flow phenomena.

\section*{Acknowledgements}
We thank G.~E. Crooks, P. Danielewicz, C. Jarzynski, M. Murray for 
helpful discussions. This work was
supported by the U.S. National Science Foundation under grant
No. PHY-0970104.



\appendix

\section{Derivation of the density of states formula}
\label{appe}

The number of states with energy less than $E$ is
\begin{equation}
\label{equ:wds26}
\Phi(E;V)=\frac{V^N}{(2\pi)^{3N}N!}\int d^{3N}p \,\theta(E-\sum_{i=1}^{N}E_i).
\end{equation}
This leads to the density of states
\begin{equation}
\label{equ:wds27}
g(E;V)=\frac{\partial \Phi}{\partial E}
=\frac{V^N}{(2\pi)^{3N}N!}\int d^{3N}p \,\delta(E-\sum_{i=1}^{N}E_i).
\end{equation}

The integral 
\begin{equation}
\label{equ:wds28}
f(E)=\int d^{3N}p \,\delta(E-\sum_{i=1}^{N}E_i)
\end{equation}
can be simplified by looking at its Laplace transform
\begin{eqnarray}
\label{equ:wds29}
F(\beta)
&=&\int_0^\infty dE\exp(-\beta E)\int d^{3N}p\,\delta(E-\sum_{i=1}^{N}E_i)
\nonumber\\
&=&\int d^{3N}p\exp(-\beta\sum_{i=1}^{N}E_i)\nonumber\\
&=&\prod_{i=1}^N\int d^3p_i\exp(-\beta E_i),
\end{eqnarray}
where $\mathrm{Re}\,\beta>0$.
For each particle,
\begin{eqnarray}
\label{equ:wds30}
\lefteqn{\int d^3p_i\exp(-\beta E_i)}\nonumber\\
\lefteqn{=4\pi\int_0^\infty 
dp_i p_i^2\exp(-\beta(\sqrt{p_i^2+m^2}-m))}\nonumber\\
\lefteqn{=\exp(\beta m)2\pi\int_m^\infty dm_\perp m_\perp^2}\nonumber\\
\lefteqn{\int_{-\infty}^\infty dy \cosh y
\exp(-\beta m_\perp \cosh y)}\nonumber\\
\lefteqn{=\exp(\beta m)4\pi\int_m^\infty dm_\perp m_\perp^2 
K_1(\beta m_\perp)}\nonumber\\
\lefteqn{=\exp(\beta m)\frac{4\pi m^2}{\beta}K_2(\beta m),}
\end{eqnarray}
where $K_1(x)$ and $K_2(x)$ are modified Bessel functions.
Therefore, 
\begin{equation}
\label{equ:wds31}
F(\beta)=\exp(N\beta m) \left(\frac{4\pi m^2}{\beta}K_2(\beta m)\right)^N.
\end{equation}

Now $f(E)$ can be calculated from $F(\beta)$ by using the inverse
Laplace transform.
\begin{eqnarray}
\label{equ:wds32}
\lefteqn{f(E)=\frac{1}{2\pi i}\int_{\beta'-i\infty}^{\beta'+i\infty}
F(\beta)\exp(\beta E)d\beta}\nonumber\\
\lefteqn{=\frac{1}{2\pi i}\int_{\beta'-i\infty}^{\beta'+i\infty}
\exp(N\beta m) \left(\frac{4\pi m^2}{\beta}K_2(\beta m)\right)^N
\exp(\beta E)d\beta,}\nonumber\\
\lefteqn{}
\end{eqnarray}
where $\beta'>0$.
The modified Bessel function can be expressed as
\begin{equation}
\label{equ:wds33}
K_2(\beta m)=\int_0^\infty dt \cosh(2t) \exp(-\beta m\cosh t).
\end{equation}
This leads to
\begin{eqnarray}
\label{equ:wds34}
\lefteqn{f(E)=(4\pi m^2)^N\prod_{j=1}^N\int_0^\infty dt_j\cosh(2t_j)}\nonumber\\
\lefteqn{\frac{1}{2\pi i}\int_{\beta'-i\infty}^{\beta'+i\infty}
\frac{\exp(\beta(E+Nm-m\sum_{k=1}^N\cosh t_k))}{\beta^N}d\beta}\nonumber\\
\lefteqn{=(4\pi m^2)^N\prod_{j=1}^N\int_0^\infty dt_j\cosh(2t_j)}\nonumber\\
\lefteqn{\frac{(E+Nm-m\sum_{k=1}^N\cosh t_k)^{N-1}}{(N-1)!}
\theta(E+Nm-m\sum_{l=1}^N\cosh t_l)}\nonumber\\
\lefteqn{=\frac{(4\pi m^3)^N}{(N-1)!m}\prod_{j=1}^N\int_0^{t^*}dt_j\cosh(2t_j)}
\nonumber\\
\lefteqn{\left(\frac{E}{m}+N-\sum_{k=1}^N\cosh t_k\right)^{N-1}
\theta\left(\frac{E}{m}+N-\sum_{l=1}^N\cosh t_l\right),}
\end{eqnarray}
where the upper bound $t^*$ is the positive solution of
\begin{equation}
\label{equ:wds35}
\frac{E}{m}+1-\cosh t^*=0.
\end{equation}










\end{document}